# Room temperature all-silicon photonic crystal nanocavity light emitting diode at sub-bandgap wavelengths

A. Shakoor,[1*] R. Lo Savio,[2] P. Cardile,[3] S. L. Portalupi,[2] D. Gerace,[2] K. Welna,[1] S. Boninelli,[3] G. Franzò,[3] F. Priolo,[3] T. F. Krauss,[1] M. Galli,[2] and L. O'Faolain[1]

[1] *SUPA*, School of Physics and Astronomy, University of St. Andrews, Fife KY16 9SS, St. Andrews, United Kingdom

[2] Dipartimento di Fisica, Università di Pavia, via Bassi 6, 27100 Pavia, Italy

[3] CNR-IMM MATIS and Dipartimento di Fisica e Astronomia, Università di Catania, via S. Sofia 64, 95123 Catania, Italy

* as829@st-andrews.ac.uk

Phone: +44-1334463091

**ABSTRACT**

Silicon is now firmly established as a high performance photonic material. Its only weakness is the lack of a native electrically driven light emitter that operates CW at room temperature, exhibits a narrow linewidth in the technologically important 1300-1600 nm wavelength window, is small and operates with low power consumption. Here, an electrically pumped all-silicon nano light source around 1300-1600 nm range is demonstrated at room temperature. Using hydrogen plasma treatment, nano-scale optically active defects are introduced into silicon, which then feed the photonic crystal nanocavity to enahnce the electrically driven emission in a device via Purcell effect. A narrow ($\Delta\lambda$ = 0.5 nm) emission line at 1515 nm wavelength with a power density of 0.4 mW/cm$^2$ is observed, which represents the highest spectral power density ever reported from any silicon emitter. A number of possible improvements are also discussed, that make this scheme a very promising light source for optical interconnects and other important silicon photonics applications.

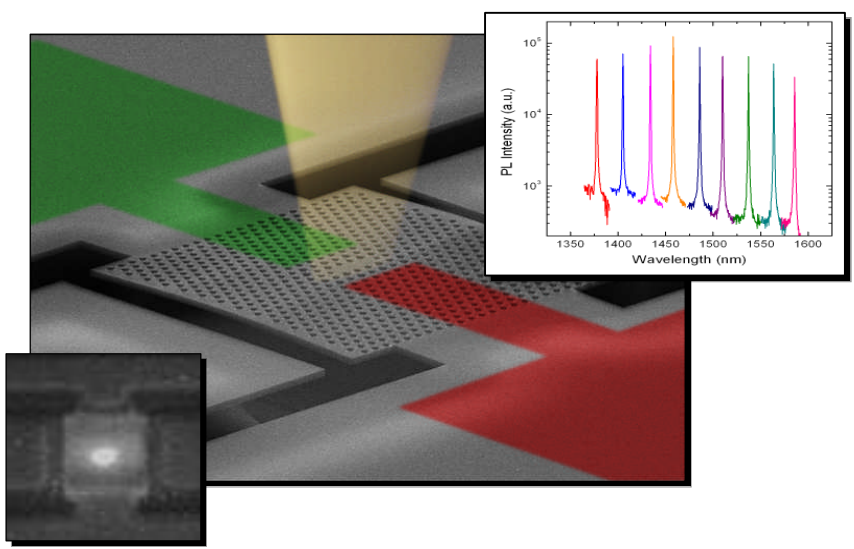

Abstract Figure: Tunable electrically pumped silicon nanolight source at telecommunication wavelengths.

# 1. Introduction

The absence of a cheap, efficient and electrically driven silicon based light emitter creates a significant barrier to the use of silicon photonics in low cost, high volume applications- a major issue for key fields such as optical interconnects and bio-sensing. Furthermore, micro- and nano- emitters are necessary for achieving a high integration density and a high channel count. The addition of III-V materials through bonding or epitaxy is currently the preferred solution giving efficient on-chip lasers [1], but the use of a costly material with complex processing makes mass manufacture challenging. A true group-IV nano light emitter remains the ultimate solution that would allow the full potential of silicon photonics to be realized. Many approaches have been employed to improve the luminescence of silicon, such as silicon nanocrystals that exhibit electroluminescence [2] and optically pumped gain [3] in the visible and near-infrared wavelength ranges. Stimulated emission has been observed from silicon quantum wells [4] and the incorporation of rare earth dopants has enabled optically pumped transparency, e.g. from erbium doped silicon nitride nano-cavities [5]. Highly doped, strained Germanium can be made to exhibit a direct transition with optically pumped lasing [6] and electroluminescence [7] reported. Raman lasers have been realised that produce very high output powers [8], however, there is no possibility of electrical driving. While these approaches have much potential, they do not combine all of the desired characteristics of a silicon light source, namely: electrical pumping, operation at sub-bandgap wavelengths, room temperature operation, small size and narrow emission linewidth.

Optically active defects offer an alternative and particularly powerful approach to improving the luminescence of native silicon. Such defects add in conserving momentum between recombining carriers, problematic due to the indirect bandgap of silicon, and may create luminescence lines and bands including emission in the important telecommunication windows [9-11]. Examples for such defect-based emission includes the formation of dislocation loops that have been used in broad area electrically driven LEDs operating near the silicon band-edge [12] or the creation of "A-centres" that have been used to demonstrate stimulated emission at cryogenic temperatures [13]. Here, we use hydrogen plasma treatment to incorporate nano-scale optically active defects into the silicon host and the high Purcell enhancement available with high Q-photonic crystal cavities to create an electrically driven all-silicon nano-LED, schematically shown in Fig. 1a, that demonstrates the desired characteristics, listed above, of a silicon light source.

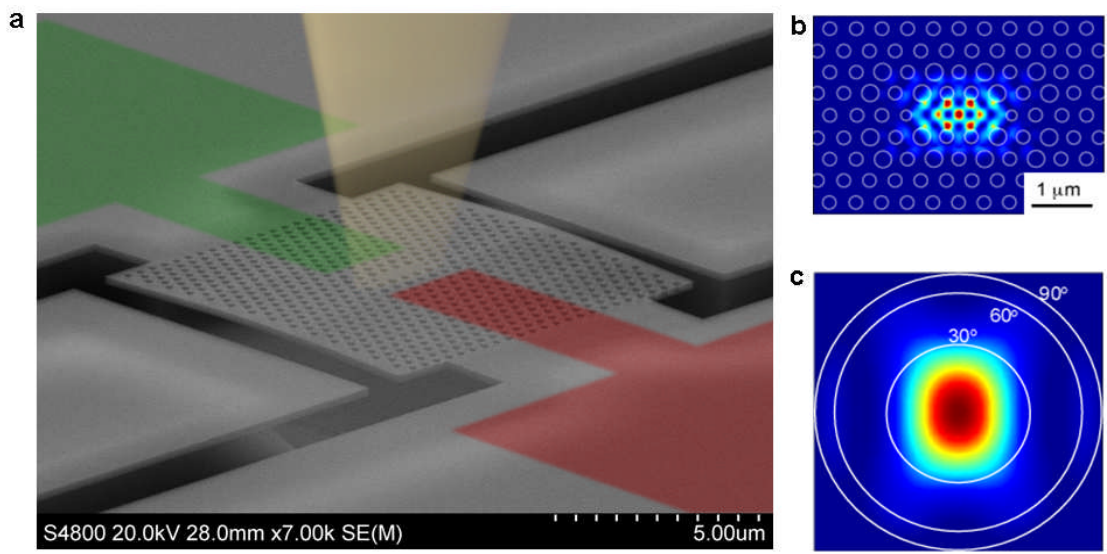

**Figure 1. Schematic and mode profile of fabricated all- silicon PhC nanocavity LED**. **a**, SEM micrograph of a silicon photonic crystal nano-LED. The doped regions are shown schematically and extend well into the photonic crystal. Far-field optimization gives a vertical, near Gaussian output beam. **b**, near-field and **c**, far-field intensity of the fundamental cavity mode, calculated by FDTD simulations of the fabricated device. A schematic view of the modified holes around the cavity is given in the near-field plot **b**.

## 2. Photonic crystal nanocavity design

We use L3 photonic crystal (PhC) cavities in a suspended 220nm thick silicon slab, fabricated using electron beam lithography and reactive ion etching. The period was 420 nm and the normalized radius (r/a) was 0.28. We have previously observed Q-values in excess of 100,000 in this system [14]. We then apply the band-folding technique to improve the vertical light extraction by engineering a near Gaussian far-field emission of the fundamental cavity mode [15, 16]. Corresponding finite-difference time-domain (FDTD) simulations are shown in Fig. 1b-c. The details of the design and fabrication of PhC nanocavity used are given in Methods section.

## 3. Creating optically active defects

Our method of creating optically active defects proceeds by treating the samples in a hydrogen plasma [9, 10], which generates surface defects at which hydrogen may get trapped [11, 17-18] or causes hydrogen to attach to pre-existing defects, i.e. created during reactive ion-etching of the PhC. To better understand the nature and the location of the defects created by the hydrogen plasma treatment, we carried out transmission electron microscopy (TEM) analysis of the hydrogen plasma treated Photonic Crystals and Czochralski silicon (Cz-Si). We performed both planar view (PV) and cross-sectional (CS) TEM measurements using a JEOL JEM 2010 instrument operating at an acceleration voltage of 200 kV.

Fig. 2a shows a PV TEM image of the region containing the PhC using bright field imaging. The black spots indicate the presence of extended defects induced by the hydrogen plasma treatment during reactive ion etching (RIE). Even though the defects are present throughout the entire silicon surface, it is evident that their concentration increases towards the sidewalls of the holes. This effect is a consequence of the plasma treatment, during which, hydrogen impacts on all exposed surfaces and produces extended damage. To better illustrate the structure of the resulting defect population, a defocused off-Bragg CS view of the plasma treated Cz-Si substrate is shown in Fig. 2b.

We notice different types of defects as a function of penetration depth. Near the surface, concentrated within the first 10 nm, the defect population is dominated by nanobubbles, whose size is a few nm. Their exact nature is unknown, but they most likely consist of agglomerates of vacancies. Going further down, we find a preponderance of platelets (indicated by white arrows in Fig. 2b), whose mean diameter is about 10-15 nm, occupying the (100) plane (parallel to the surface) and the {111} planes. Similar platelets were observed in refs [19-21]. A high resolution image of a few platelets is shown in the top left inset in Fig. 2b. Moreover, some of the dark traces, located between the two previously described regions, exhibit the typical "coffee bean" shape indicative of dislocation loops. One of them is shown in the top right inset of Fig. 2b. While the existence of these hydrogen plasma induced defects has been described before [19-21], their luminescence properties are still not well understood.

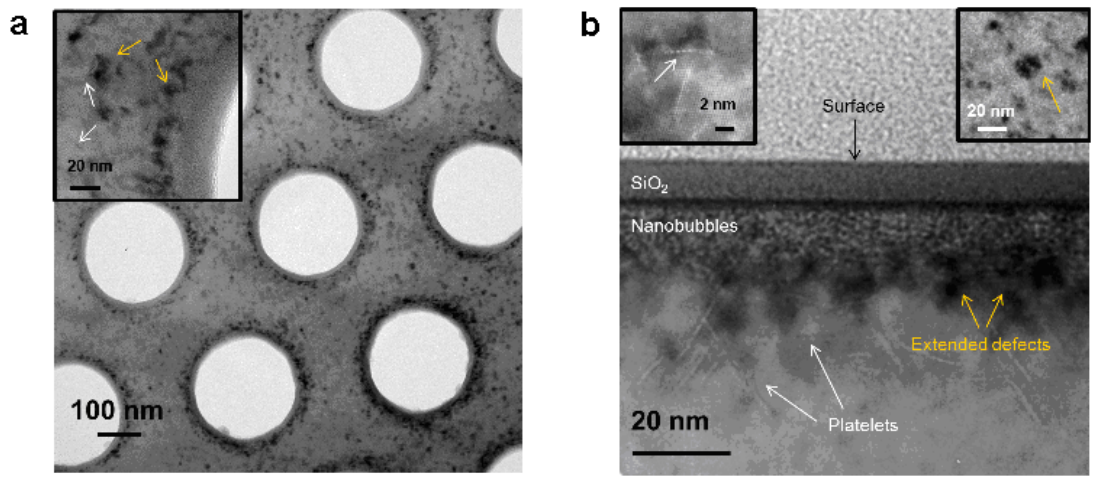

**Figure 2. Transmission electron microscopy (TEM) images showing the defects in silicon created by hydrogen plasma treatment**. **a**, Planar view (PV) TEM image of hydrogen plasma treated PhC. The defects concentration increases in correspondence of the holes sidewalls. The inset shows a zoom-in region close to a single hole, where some defects are indicated by arrows. **b**, Cross- sectional (CS) view of a plasma treated Cz-Si sample. White arrows indicate platelets and yellow arrows indicate extended defects. In the top left inset a high resolution image of the platelets is given, while in the top right inset a typical coffee bean shape of a dislocation loop is shown.

### 4. Luminescence of optically active defects within a PhC nanocavity

Fig.3 shows the photoluminescence (PL) spectra of hydrogen plasma treated PhC nanocavities with different lattice constants, as compared with that of bulk silicon (Cz-Si) (see Methods section for details of the excitation scheme). The photoluminescence signal from the bulk silicon is very weak, while the treated PhC cavities show a background signal with strong, sharp peaks corresponding to the fundamental mode of each PhC cavity. The enhancement of background PL is due to incorporation of optically active defects by hydrogen plasma treatment while the peak reflects the enhancement due to the Purcell effect and the improved extraction due to the nanocavity resonance and Photonic bandgap effects. Combining these effects, we observe an overall 40000-fold (4 orders of magnitude) increase of the PL signal at room temperature relative to bulk silicon. As demonstrated in Fig 3, the emission line is easily tunable through the 1300-1600 nm wavelength range and also demonstrates the robustness and repeatability of this method. Following refs. 22-24, we estimate a Purcell enhancement factor of ~10, which is responsible for an increase in the radiative recombination rate increase and subsequent suppression of thermal quenching.

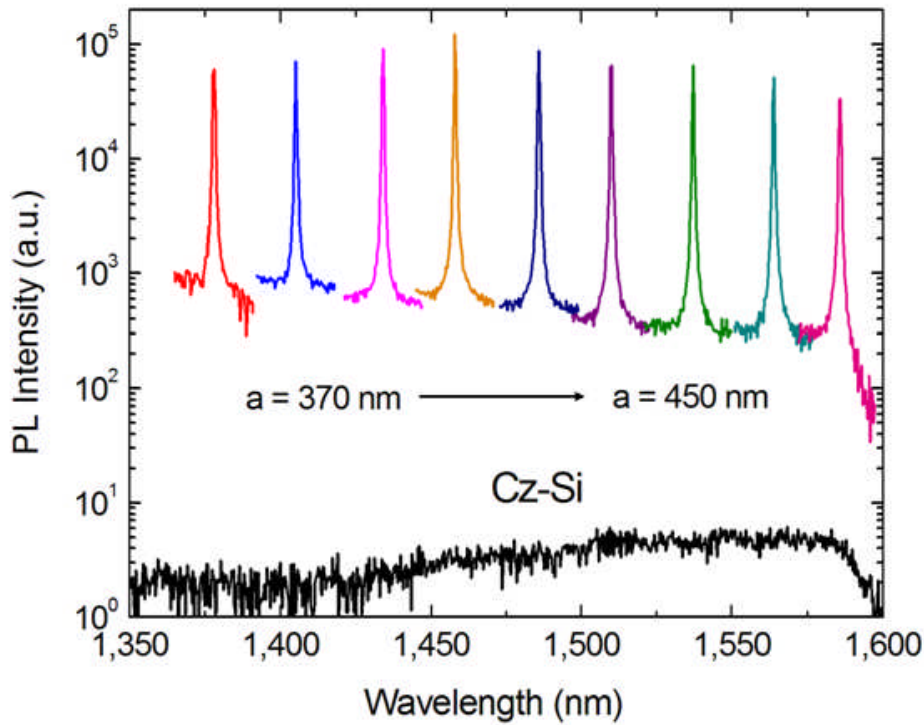

**Figure 3. Comparison of the photoluminescence of hydrogen treated nanocavities with Czochralski silicon.** Peaks correspond to the fundamental mode of cavities with different lattice period, a. The emission line (PL of fundamental cavity mode) of the hydrogen plasma treated cavity is over four orders of magnitude higher than bulk silicon over the entire range and tunable between 1300 and 1600 nm.

We found that the key parameter is the duration of the plasma treatment, which impacts both on the total PL intensity and on the cavity quality factor. The PL emission is characterized by two broad bands centred around 1300 and 1500 nm (see Fig. 4a). In Fig. 4b we show the PL emission intensity, integrated between 1400 and 1600 nm, for different treatment duration and for two different plasma compositions - pure hydrogen (20 sccm gas flow) and a hydrogen/argon mix (20 sccm and 4sccm gas flow for hydrogen and argon respectively). The signal from the samples treated with pure hydrogen is clearly stronger and quickly reaches the maximum value. The argon was added to the plasma in order to elucidate the role of physical damage, which turns out to inhibit the signal enhancement, especially for the longer treatment times; using pure argon plasma gave a very weak PL signal (not shown), clearly suggesting that hydrogen plays the key role in the emission process [19].

In addition, we also measured the cavity Q-factors using the resonant scattering technique [14] and a typical spectrum is shown in Fig. 4c. The Q-factor follows a similar trend as the PL intensity for different treatment duration, Fig. 4d, that is, an increase for hydrogen-treated cavities, and a decrease when the plasma contained argon. This is also consistent with our observation that argon causes physical damage and introduces roughness to the silicon surface. Such roughness increases the optical loss of the PhC cavity, thus reducing the overall Q-factor. In contrast, the hydrogen plasma treatment increased the Q-values. In fact, a close comparison of SEM pictures, not shown here, indicates that the sidewalls of the PhC cavities become smoother after hydrogen plasma treatment, giving rise to a smoothing or "plasma polishing" effect that can explain the observed increase in Q-value. Overall, we conclude that hydrogen has the dual benefit of increasing both the PL signal and the Q-factor.

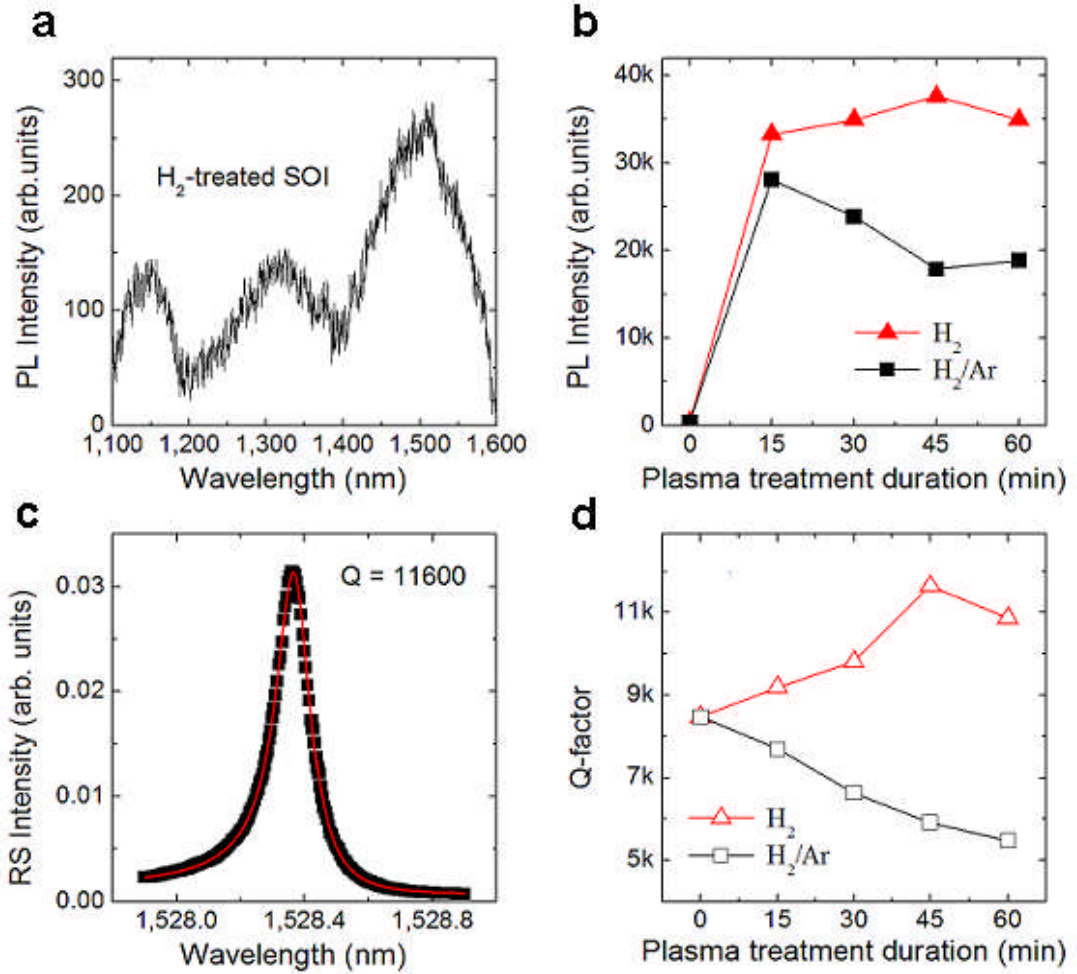

**Figure 4. The effect of plasma treatment duration on photoluminescence and the cavity Q-factor. a**, Photoluminescence observed from hydrogen-treated silicon-on-insulator (SOI). **b**, Effect of treatment duration on the integrated PL intensity; the red curve represents treatment with pure hydrogen plasma and the black curve shows treatment with a mixed hydrogen/argon plasma. **c**, Resonant scattering spectrum of the highest Q-cavity with the best-fit of the experimental data to a Fano lineshape (red line) [14]. **d**, Effect of different plasma conditions on the cavity Q factors.

These results were achieved for samples treated after the photonic crystal etch step, i.e. in the presence of the etched holes as shown in Fig. 2a. We also performed the plasma treatment prior to the etching step and found a much weaker enhancement.

This reduced enhancement highlights the importance of the exposed surface area and that the observed PL enhancement is mainly a surface effect. Clearly, the hydrogen ions only have very low energy (≈ 400 eV) and, consequently, a low penetrating power into the bulk material. In addition, hydrogen plasma treatment is known to passivate the silicon surface. We expect that it reduces carrier recombination at surface defects and thus improves the efficiency of radiative recombination [25].

### 5. All-silicon nano- LED

Finally, we proceed to electroluminescent devices. Since photonic crystal fabrication is fully compatible with ULSI processes [26], we create *pin* junctions using multiple lithography and ion implantation steps providing a monolithic source of electron-hole pairs to feed our optically active defects. Here, we create fingers of doped regions, as marked in Fig. 1a, which extend into the photonic crystal, similar to [27]. This forces carriers to recombine in the cavity by virtue of the low resistivity of the two highly doped fingers. The conditions were carefully optimised to provide the best possible current injection, and to finely control the position of the depletion region with respect to the cavity region.

Electroluminescence (EL) was generated by applying a forward bias across the junction. Light was collected with the same experimental apparatus used for PL measurements, thus allowing a direct comparison between PL and EL emission intensities. This comparison is shown in Fig. 5a, where we report the maximum power spectral density expressed in pW/nm for the fundamental cavity mode for both PL and EL. The photoluminescence is recorded for a pump power of 0.8 mW and at an excitation wavelength of 640 nm. The electroluminescence is recorded at an applied voltage of 3.5 V, with a current of 156.5 µA, thus consuming an electrical power of 0.55 mW. Remarkably, the EL signal is more intense than the PL signal across the entire spectral range, in contrast to that usually observed, and is a testament to the potential of this system as an electrically driven source. The PL signal is lower than EL for almost same input power due to the low absorption (~5%) of the thin silicon slab at the 640 nm excitation wavelength.

Fig. 5 also shows an optical image of the device under zero voltage (Fig. 5b) and with a 3.5 V bias applied (Fig. 5c). The latter image was captured with an infra-red (IR) camera and the bright emission spot is clearly visible as soon as the voltage is turned on. The Q-factor of the cavity for the electrical devices was 4000, while for bare PhC cavities (before device fabrication) higher Q-values were observed (see Fig. 4c). This reduction is a consequence of imperfections introduced during the *pin* junction fabrication process as well as free carrier absorption.

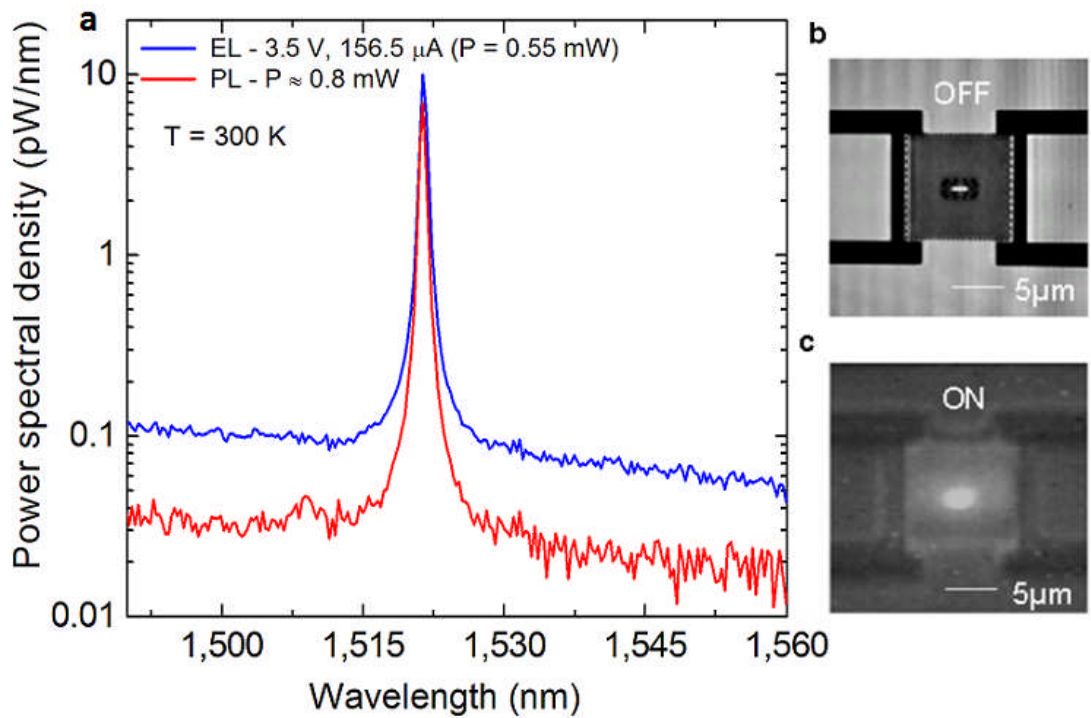

**Figure 5: Electroluminescence and Photoluminescence from the device**. **a**, Comparison of EL and PL from a PhC nanocavity (with integrated pin junction) treated with hydrogen plasma. **b**, Micrograph (top view) and **c**, filtered IR picture of the device showing strong electrically driven emission from the silicon PhC

nanocavity at room temperature (a low-pass filter with a cut-off at 1500 nm was used). Integrating over the fundamental cavity mode, the emission power is 4pW (electrically driven). Integrating over the full wavelength range (1200-1600nm), we measure a total power of 45 pW. The actual power generated in the cavities is 10-100 times higher than the measured output powers, considering the percentage of light emitted vertically, collection efficiencies and losses in the setup. It should also be noted that only a fraction of the optical pump is absorbed in the top silicon layer, resulting in coincidentally similar values for the optical and electrical pump power.

The spectral density of the emission is the key figure of merit for many applications such as Wavelength Division Multiplexing (WDM) for communications, multimodal operation for biosensing or spectral coherence for interferometry. In the fundamental cavity mode, our device gives a spectral density of 10 pW/nm (by considering the active area, 800μW/nm/cm$^2$), which is by far the highest reported value for any silicon-based electrically-driven nano-emitter (even without restricting the comparison to room temperature emission and telecom wavelengths- see table 1), thus encouraging further efforts for the realization of the first Si-based electrically pumped laser. The compatibility of these optically active defects with electrical driving even allows the traditionally inefficient silicon emission to give a measured output power that is comparable with that of most electrically driven III-V photonic crystal-lasers and LEDs [28-30], where collected powers are reported to be in the picowatt to nanowatt level, similar to the power levels reported here. Additionally, our nano-LED is easily tunable in the 1300-1600 nm range, by chnging the lattice period of the PhC lattice, as demonstrated in Fig 3.

Table 1. Comparison of different band-edge and sub-bandgap silicon light sources

| | Injection Current | Operating wavelength λ (nm) | Operating temperature (K) | Emission linewidth Δλ (nm) | Power density* μW/cm$^2$ | Power spectral density μW/nm/cm$^2$ | Wall plug efficiency** | Comments |
|---|---|---|---|---|---|---|---|---|
| Si:Dislocation loops [12] | 50mA | 1150 | ~300 | 90 | 600 | 6.6 | 10$^{-4}$ | Broad area device, |
| Band-edge (structured) [31] | 130mA | 1150 | ~300 | 50 | 300*** | 6*** | 10$^{-2}$ | Broad area device, |
| A-centers [13] | optical | 1280 | <70 | 0.5 | a.u | a.u | not reported | Works only at cryogenic temperatures. |
| W-centres [32] | 2mA | 1218 | 6 | ~1 | 2.7 | 2.7 | 10$^{-7}$ | Works only at cryogenic temperatures |
| Erbium in SiN (membraned) [5] | optical | 1565 | 5.5 | 0.03 | a.u | a.u | not reported | Works only at cryogenic temperatures. |
| Erbium in SiN (on Si) [33] | 1.5A/cm$^2$ Optical | visible 1550 | ~300 | 100 | a.u | a.u | not reported | |
| Bandedge (PhCs) [24,34] | optical | 1100 | ~300 | <0.5 | a.u | a.u | not reported | Band-edge emission |
| This work | 156 μA | 1200 to 1600 | ~300 | 0.5 | 400 | 800 | 10$^{-8}$ | Sub-bandgap, room temperature operation, electrically pumped, small size. Suitable for interconnect and other silicon photonics applications |

\* Calculated using the emitting area

\*\* The wall plug efficiency is calculated by considering the measured output power.

\*\*\* Output powers expressed in arbitrary units (a.u). Power calculated on the basis of efficiencies and drive current, voltage and area.

As a consequence of the very narrow emission linewidth (0.5 nm), the power efficiency of our LED is $0.7 \times 10^{-8}$, (based on the collected power). This efficiency should hence be compared with those of other micron scale emitters with narrow linewidths. In fact, our efficiencies are within an order of magnitude of those reported- $1.6\text{-}3 \times 10^{-8}$ [29] and $8 \times 10^{-8}$ [28]- for direct bandgap (III-V) lasers operating below threshold, a remarkable result considering the indirect band gap of silicon.

The device is temperature stable up to $350^{\circ}C$, after which hydrogen starts to diffuse out of the silicon matrix, resulting in a reduction of the luminescence signal. For example, annealing the device at $500^{\circ}C$ for 1 hour reduced the luminescence by a factor of 4 to 5. This also indicates the importance of hydrogen in the emission process. The physical defects created by the plasma treatment and decorated by hydrogen are therefore essential for the emission process.

There is a very slight (0.1nm) red shift of the mode under active bias conditions, There may also be a masked blue shift due to the injected carriers, estimated to be less than 0.5nm [22,27]. This indicates that the temperature of the device increases by only $10\text{-}20^{\circ}C$, which is well below the temperature beyond which hydrogen outdiffusion occurs (please see methods section for more details). Damage to the device or a reduction in the EL level was also not observed under active bias conditions. Thus no detrimental effect of the active bias conditions was apparent during this experiment.

There is still considerable scope for improvement of our device, and it is realistic to expect laser operation in due course. A key to improving the operation and device efficiency is to better understand the incorporation of defects into the silicon host. The low ion energies in the plasma treatment we currently use results in low penetration depths of the ions, resulting in defect formation close to the surface (see Fig. 2b), which limits the active volume available. The luminescence is also very broadband indicating that most of the emitters are off resonance and do not contribute to the luminescence; therefore, only a tiny fraction of the defects actually emits into the cavity mode and experience PhC enhancement. Ion implantation, the preferred approach for defect based photodetection [35], has much potential in this respect, as it gives access to a wide range of species and energies thus providing a large parameter space for further optimization. The small mode volume of our device, though important for a high Purcell factor, is also a big limitation with respect to total output power. This can be addressed by using a coupled cavity configuration which is a promising route to combining high Purcell factors and large mode volumes [36, 37]. Using different cavity designs, such as the H0-type [38], will also ensure single mode operation across the entire gain bandwidth.

We estimate that by maximizing the number of defects in the active volume, more than a factor of 10 increase in luminescence is possible. In addition, a better understanding of the nature of the defects may lead to a narrower emission bandwidth, potentially adding another order of magnitude. This will also lead to dramatic improvements in the efficiency, as in the current device, due to the low number of defects, the bulk of the current passes through the device without recombining. A successful combination of these effects thus makes our approach a very promising route towards an all-silicon electrically driven nano-laser.

## 6. Conclusions

In conclusion, we have enhanced luminescence from silicon by more than 4 orders of magnitude by combination of hydrogen plasma treatment (up to 2 orders of magnitude), and both Purcell and extraction efficiency enhancement (300-400 times), and demonstrated an all-silicon nano-LED with the highest power spectral density reported in silicon to date (see table 1). The Photonic Crystal cavity suppresses thermal quenching of the hydrogen plasma induced optically active defects and improves the emission rate, thereby making them a route to a practical light source. Our nano-LED operates at room temperature and in the technologically important wavelength range of 1300-1600 nm. As our approach can provide narrow linewidth nano-LED anywhere in this range, it is ideally suited to Wavelength Division Multiplexing (WDM) like applications. Additionally, it provides a means to realising high spectral purity light sources in a CMOS environment, key to making practical devices such as inexpensive biosensors.

**Methods**

**1. Design and fabrication of Photonic crystal nanocavity**

L3 PhC nanocavities were realized by removing three holes from the Γ-K direction of a PhC pattern, characterized by a period of a = 420 nm and a normalized radius (r/a) = 0.28. The two holes adjacent to the cavity were reduced in size and displaced laterally to increase the Q-value of the cavity [39]. These parameters produce an L3 nanocavity with a fundamental mode around 1.5 µm wavelength. To obtain the maximum out-coupling efficiency in the vertical direction, we applied a far-field optimization technique whereby alternating holes around the cavity are enlarged to form a "second order grating", which significantly increases the vertical out-coupling efficiency [16]. Here, we used an enlargement of 15nm in hole radius to provide maximal vertical out-coupling efficiency. The far-field optimization reduces the Q-values of the cavity, but, similar to [22], the Q-values are, in fact, limited by free carrier absorption. A range of PhC cavities were realized with period spanning from 370 to 450 nm in order to demonstrate that we are able to very precisely control the fundamental cavity mode in the range 1300 - 1600 nm.

The fabrication of L3 PhC nanocavities was carried out by electron beam lithography and reactive ion etching. We used $CHF_3/SF_6$ gas chemistry for etching the PhC holes. A free standing membrane was formed by removing the supporting silicon oxide layer using wet chemical etching by hydrofluoric acid.

Importantly, for large scale integration, undercutting is not necessarily a prerequisite for high quality Photonic Crystals [40]. Oxide cladding may instead be employed providing a very stable and robust device. Here, we use membraned devices as they provide the greatest operating tolerances, which is useful for initial demonstrations.

**2. Ion implantation for *pin* junction**

The n-type finger-like arm of the device was doped by ion implantation to a level of $10^{19}$ P/cm$^3$ and it was separated from the corresponding p-type finger (doped with $10^{19}$ B/cm$^3$) by a distance of 500 nm. This separating region is slightly p-type doped with ~$10^{15}$ B/cm$^3$ (background doping of the as-bought SOI wafer). Alignment between the steps is carefully carried out by means of electron beam lithography. For these conditions, the cavity region is fully depleted. Therefore, the intrinsic region is filled with carriers during the operation that then recombine in an efficient way. Doped silicon is known to be a source of optical losses; however, doping has been shown to be relatively insignificant in this system with Q-factors up to 40,000 observed for carrier densities of $10^{18}$/cm$^3$ [41].

**3. Plasma treatment**

The plasma treatment was performed in a parallel plate reactive ion etching system with a hydrogen and hydrogen/argon flow of 20 sccm and 24 sccm respectively, a pressure of $1\times10^{-1}$ mbar, an RF power of 40 W, resulting in a DC bias of ≈ -400 V. This treatment step was carried out on the membraned PhCs after dopant and contact annealing was performed. Importantly, since the hydrogen plasma treatment is carried out as a post process, it is compatible with silicon photonics device processing and is stable under operating temperature of silicon photonic devices. Furthermore, the defects created by the treatment were found to be reasonably stable, with only a moderate decrease in the emission observed over 6 months.

**4. Characterization**

For characterization, we used room temperature confocal PL and the resonant scattering method to measure the PL and Q-values of the nanocavity modes, respectively. The cavities were excited with a CW diode laser emitting at 640nm. The spot was focused to 1 µm$^2$ in the centre of a cavity with a microscope objective (NA= 0.8) and the emitted light was collected back with the same objective and fed to a grating spectrometer. Further details of these characterization methods are given in [14, 22].

The emitted light from the electrically driven device was collected using the same PL setup. From Fig. 5, it is observed that the EL signal (also CW driving) has more emission power compared to PL across the entire spectral range. This indicates that the injection of carriers is relatively efficient in this scheme. The PL signal is lower than EL for almost same input power due to the low absorption (~5%) of the thin silicon slab for the 640nm excitation wavelength.

In addition, we notice that the device heating is similar under both optical and electrical pumping, and is estimated to be of the order of 1-2°C from the small redshift (about 0.1 nm) of the cavity resonance as compared to very low power excitation. This is consistent with previous literature works on silicon and GaAs membrane PhC nanocavities under similar pumping conditions [42-43]. In fact, despite the very high power density in the cavity region under both electrical and optical pumping (about 50 kW/cm$^2$), a small temperature increase is observed due to the very efficient heat dissipation occurring in membraned PhC nanocavities [43]. Oxide cladding, see above, is expected to further improve heat dissipation.

**Authors Contributions**

AS and LOF conceived the idea of plasma treatment of PhC cavities. AS developed the plasma treatment process (under the guidance of TFK). AS, PC, SLP and LOF designed and fabricated the samples. RLS and MG carried out all the optical measurements. PC and GF performed the ion implantation and SB performed the TEM analysis, guided by FP. DG and KW designed and modelled the PhC cavities. All authors contributed to discussions and analysis of data. AS, PC, MG and LOF wrote the manuscript, with contributions from the other co-authors. TFK, MG, LOF and FP coordinated and directed the work.

**Acknowledgements**

We acknowledge Lucio C. Andreani (Dept. of Physics, Univ. of Pavia) for useful discussions and suggestions, and A. Liscidini (Dept. of Electronics, Univ. of Pavia) for providing wire bonding to the EL devices. This work was supported by Era-NET NanoSci LECSIN project coordinated by F. Priolo, by the Italian Ministry of University and Research, FIRB contract no. RBAP06L4S5 and the UK EPSRC UK Silicon Photonics project. The fabrication was carried out in the framework of NanoPiX (see http://www.nanophotonics.eu).